\documentclass[%
 reprint,
 amsmath,amssymb,
 aps,
]{revtex4-1}

\usepackage{graphicx}
\usepackage{dcolumn}
\usepackage{bm}
\usepackage{extarrows}
\usepackage{amssymb}

\begin{document}

\preprint{APS/123-QED}

\title{Perfect state transfer and efficient quantum routing: a discrete-time quantum walk approach}

\author{Xiang Zhan}
\affiliation{Department of Physics, Southeast University, Nanjing 211189, China}
\author{Hao Qin}
\affiliation{Department of Physics, Southeast University, Nanjing 211189, China}
\author{Zhi-hao Bian}
\affiliation{Department of Physics, Southeast University, Nanjing 211189, China}
\author{Jian Li}
\affiliation{Department of Physics, Southeast University, Nanjing
211189, China}
\author{Peng Xue}
\email{gnep.eux@gamil.com}
\affiliation{Department of Physics, Southeast University, Nanjing 211189, China}

\date{\today}

\begin{abstract}
We show a perfect state transfer of an arbitrary unknown two-qubit state can be achieved via a discrete-time quantum walk with various settings of coin flippings, and extend this method to distribution of an arbitrary unknown multi-qubit entangled state between every pair of sites in the multi-dimensional network. Furthermore, we study the routing of quantum information on this network in a quantum walk architecture, which can be used as quantum information processors to communicate between separated qubits.
\begin{description}
\item[PACS numbers]
03.67.Ac, 03.67.Lx, 03.67.Bg, 03.67.Hk
\end{description}
\end{abstract}

\pacs{Valid PACS appear here}
\maketitle


\section{INTRODUCTION}
Quantum state transfer ~\cite{Bose,DTQWST,networkST,Christandl,Kendon,Chudzicki} between arbitrary distant sites is an important task for quantum information processing. The perfect state transfer can be implemented via quantum data bus to connect distant sites on a quantum network. In this paper we consider two related research fields, quantum walks (QWs) on one-dimensional or higher-dimensional lattices and perfect state transfer. We present a realization of arbitrary unknown state transfer in a QW architecture.

Quantum walks were first introduced by Aharonov et. al.~\cite{QW} in 1993. Their work has shown that the average path length, which corresponds to the propagate velocity of the walker, can be much larger than that of the classical random walk, due to quantum interference effects. This property makes QW to be promising to develop new quantum algorithms~\cite{Kempe}. The QW is also an promising resource for quantum simulation~\cite{Kitagawa,X1,X2,np,Oliveira,burnett,segwa} and for universal quantum computation~\cite{Childs,Childs2,Lovett}. Various studies for implementing QWs in different physical systems have been made~\cite{Travaglione,Sanders,Dur,X09,X5}. Experimental realizations in various physical systems have been reported~\cite{Du,Perets,Karski,Schreiber,Bouwmeester,Zahringer,Schmitz,Cote,Do,Zhang,Peruzzo,Broome,Schreiber2,Sansoni}.

The QW has been proposed as an efficient way to perfectly transfer unknown quantum states and therefore transfer quantum information without requirement of extra control~\cite{Bose,DTQWST,networkST,Christandl,Kendon,Chudzicki}. The task of quantum state transfer has been introduced in the context of quantum computation as a protocol to establish interactions between separated qubits on the network. The network can be used for the transfer and distribution of arbitrary unknown qubits and multi-qubits entanglement, which are important resources for teleportation~\cite{teleportation}, error correction~\cite{Brun} and the other tasks~\cite{Jozsa} in quantum information processing.

Under the evolution of QW, with full control of the walker+coin system the target state can be transferred between arbitrary input and output ports in a network of arbitrary spatial dimension. We regard the task as perfect routing which is more useful in a computational architecture than state transfer. We show how arbitrary networks can be designed for routing multi-qubit quantum states between arbitrary sites with a time scaling that is linear to the distance to be covered.

In this paper, we present a scheme on perfectly transferring unknown state and routing quantum information in architecture of  QWs on regular network. We transfer the coin states between arbitrary sites by manipulating the coin flipping. Our protocol benefits from the full control of the walker+coin system. Compared to the previous protocols on quantum state transfer, transferring the coin state is more feasible and easier to extend to multi-qubit entanglement transfer releasing the requirement of periodicity of QWs. The walker carries the coin prepared in a certain state, which needs to be transferred, and walks from the original site to the target site with time-dependent coin flippings for each step. Thus the coin state is perfectly transferred between two sites. Furthermore, we extend the method to distribute multi-qubit entanglement between arbitrary sites resulting in a possible implementation of efficient quantum routing.

This article is organized as follows. In Sec. \uppercase\expandafter{\romannumeral2}, we give a brief introduction to a discrete-time QW on line, and illustrate two kinds of biased coin flipping operators, which play a critical role in our scheme. The content of the scheme for realizing quantum routing based on controllable perfect state transfer via discrete-time QW is showed in Sec. \uppercase\expandafter{\romannumeral3}. We extend the scheme from one dimensional quantum routing to two dimensional case, routing entangled qubits to two arbitrary positions. We also show the scheme can be extended to high dimensional case.
\section{DISCRETE-TIME QW AND BIASED COIN FLIPPING OPERATORS}
One-dimensional (1D) QW on the line describes QW of a walker whose motion is restricted on the line~\cite{Kempe, line}. For the discrete-time QW, there is another freedom degree which acts as the coin. The state of the walker-coin system is denoted by a vector in the Hilbert space $\mathcal{H}=\mathcal{H}_w\otimes\mathcal{H}_c$, where the subscripts $w$, $c$ stand for walker and coin respectively. The motion of the walk is conditioned by the coin state via a conditional shift operator
\begin{equation}
\label{eq:1}
S=\sum_{x}(|x+1\rangle\langle
x|\otimes|0\rangle\langle0|+|x-1\rangle\langle x|\otimes|1\rangle\langle1|),
\end{equation}
where the summation symbol denotes for sum over all possible positions. The relations of the coin states and the directions of walk are showed in Fig. 1(a). The whole walking process is realized by repeating the sequence of the coin flipping operator and the conditional shift operator Eq. (\ref{eq:1}) step by step (so called discrete-time). The coin flipping operator is $\mathbb{I}\otimes\mathbb{C}$, where $\mathbb{I}$ is the identity operator of the walker, and $\mathbb{C}$ is the flipping operator applied on the coin state.
Standard discrete-time QW is performed with unbiased coin flipping operators (Hadamard operator). Here we introduce two biased coin flipping operators, the identity operator $\mathbb{I}$ and the Pauli operator $\sigma_x$, which are used in our scheme. The coin states remain unchanged under the coin operator $\mathbb{I}$ and therefore the direction of the walker keeps same as that for the previous step. The Pauli operator $\sigma_x$ performs exactly contrary with $\mathbb{I}$ . It makes the coin states be entirely flipped (i.e. $|0\rangle\rightarrow|1\rangle$, $|1\rangle\rightarrow|0\rangle$), and the walker turns to the direction reversed to that for the previous step. (Seeing Fig. 1 (b))
\section{PERFECT STATE TRANSFER and EFFICIENT QUANTUM ROUTING}

\subsection{Perfect state transfer via 1D discrete-time QW}
We introduce a basic scheme to transfer the coin state through 1D discrete-time QW firstly. The initial state of the walker-coin system is
\begin{equation}
\label{eq:2}
|\psi\rangle^{0}=|0\rangle(\alpha|0\rangle+\beta|1\rangle),
\end{equation}
where $|\psi\rangle^i$ denotes the state describing the system after the $i$th step, $\alpha$ and $\beta$ are complex numbers, and $|\alpha|^{2}+|\beta|^{2}=1$. In our scheme the coin states $(\alpha|0\rangle+\beta|1\rangle)$ is transferred to a certain position $x$ after arbitrary $n$ steps discrete-time QW. The total process can be described as
\begin{equation}
\label{eq:3}
|\psi\rangle^{0}=|0\rangle(\alpha|0\rangle+\beta|1\rangle)\xlongrightarrow{\text{n\, steps}}|\psi\rangle^{n}=|x\rangle(\alpha|0\rangle+\beta|1\rangle).
\end{equation}

To ensure the transfer perfect, the state of the $(n-1)$th step needs to be at the position $(x+1)$ or $(x-1)$. Because states in other positions can not walk to the position $x$ in one step, which result in the possibility of the walker in position $x$ less than 1. Thus the state after the $(n-1)$th step is
\begin{equation}
\label{eq:4}
|\psi\rangle^{n-1}=A_{1}|x-1\rangle|\phi\rangle_{(n-1),(x-1)}+A_{2}|x+1\rangle|\phi\rangle_{(n-1),(x+1)},
\end{equation}
where $A_{j}$ ($j=1$, $2$) is the complex amplitude, $|A_{1}|^{2}+|A_{2}|^{2}=1$, and $|\phi\rangle_{n,x}$ is the coin state corresponding to the walker in position $x$ after the $n$th step. Since the walker is able to propagate to $(x-1)$ or $(x+1)$ at the $(n-1)$th step, we have
\begin{equation}
\label{eq:5}
n-2\le x \le n+2,
\end{equation}
which restricts the range of the position that the coin state can be perfectly transferred. Another restriction is that for even (odd) step numbers the coin state can be only transferred to the even (odd) positions. For example, if the walking was stated from $|0\rangle$, which is
 \begin{equation}
 \label{eq:6}
 \frac{n-x}{2}\in Z,
 \end{equation}
where $n$ is the number of the step, $x$ is the position, and $Z$ stands for the combination of all integers.
Performing the $n$th step gives
\begin{align}
\label{eq:7}
|\psi\rangle^{n}&=S(\mathbb{I}\otimes \mathbb{C})|\psi\rangle^{n-1}\notag\\
                &=S(A_{1}|x-1\rangle(\mathbb{C}_{n-1,x-1}\otimes|\phi\rangle_{(n-1),(x-1)})\notag\\
                 &\quad +A_{2}|x+1\rangle(\mathbb{C}_{n-1,x+1}\otimes|\phi\rangle_{(n-1),(x+1)})),
\end{align}
where $\mathbb{C}_{n,x}$ stands for the coin flipping operator performed on the position $x$ of the $n$th step. After $(n-1)$ step, the walker in the positions $(x\pm1)$ walks to the position $x$ at the $n$th step. Thus we can obtain $\mathbb{C}_{n-1,x-1}\otimes|\phi\rangle_{(n-1),(x-1)}\rightarrow|0\rangle$, and $\mathbb{C} _{n-1,x+1}\otimes|\phi\rangle_{(n-1),(x+1)}\rightarrow|1\rangle$. Thus, $\alpha$ is dependent on $A_1$ only and $\beta$ on $A_2$ only. In our scheme the unknown information carried by $\alpha$ and $\beta$ can be transferred to the position $(x\pm1)$ after $(n-1)$ steps respectively and then after the $n$th step the state $\alpha|0\rangle+\beta|1\rangle$ is transferred to the position $x$. We show the detailed process below.
  \begin{figure}[htb]
  \includegraphics[width=8cm]{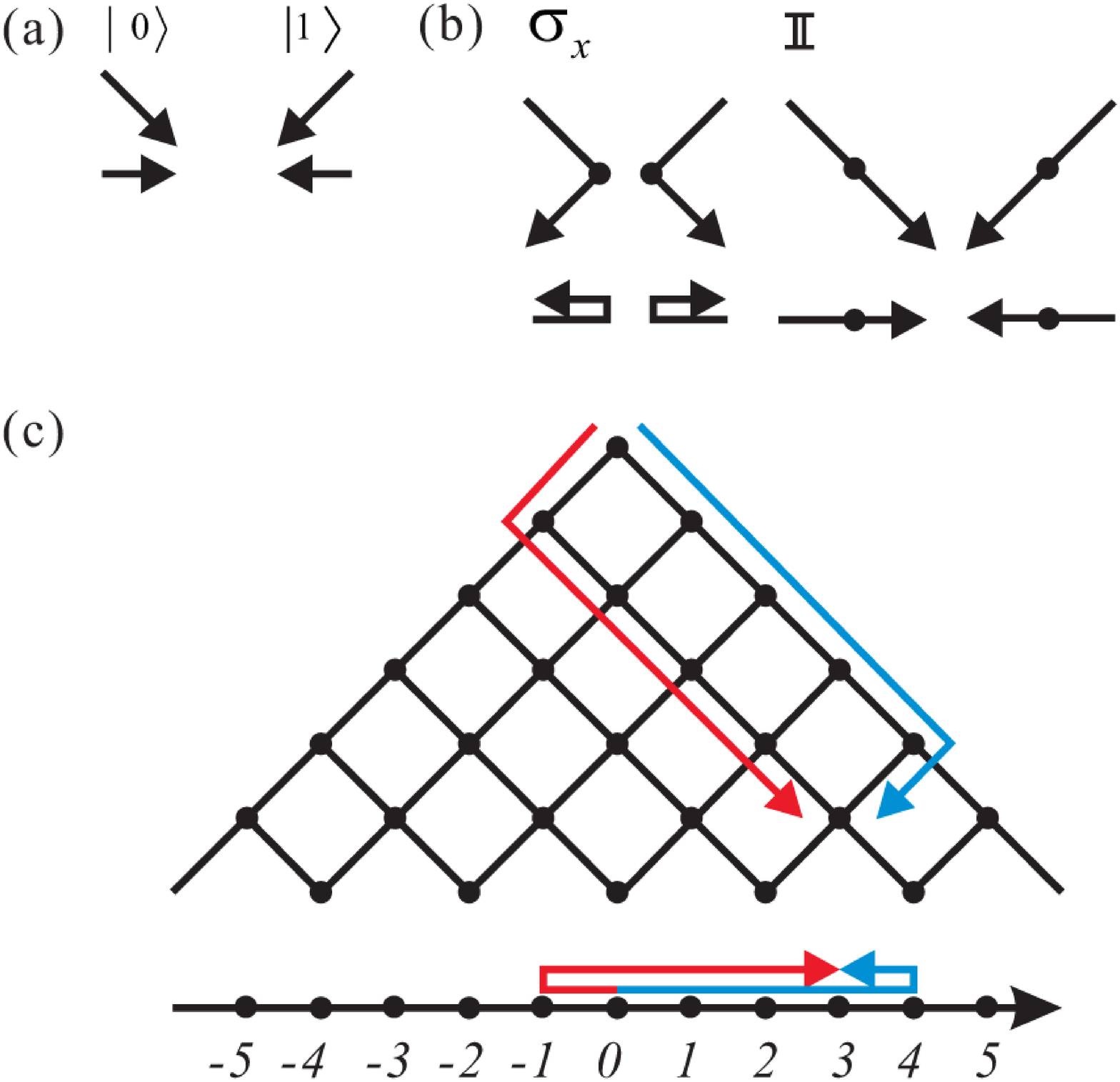}
  \caption{The scheme of state transfer in 1D discrete-time QW architecture. (a) The walking direction correspond to the resulting coin state. (b) The walking directions controlled by the biased coin flippings. (c) Perfectly transferring the coin state $\alpha|0\rangle+\beta|1\rangle$ from position $0$ to $3$ after 5 steps discrete-time QW. The red and blue arrows indicate the directions of the information flows of $\alpha$ and $\beta$ respectively. The protocol is showed in both tree diagram (the upper one) and the number axis case.}
  \label{fig:1}
  \end{figure}

The whole state of the walker-coin system after the $1$st step is $|\psi\rangle^{1}=S(\mathbb{I}\otimes \mathbb{C}_{0,0})|\psi\rangle^{0}$. We choose $\mathbb{C}_{0,0}=\sigma_x$, and then get
\begin{equation}
\label{eq:8}
|\psi\rangle^{1}=\beta|1\rangle|0\rangle+\alpha|-1\rangle|1\rangle.
\end{equation}

We now show how the walking be performed between the $1$st and the $(n-1)$th step (totally $(n-2)$ steps). We also use $\sigma_x$ or $\mathbb{I}$ as the coin flipping operators all through the process.

The walker in the position $1$ walks straight (along right) for $a$ steps and then turns around (along left) for $b$ steps to reach the position $(x+1)$ at the $(n-1)$th step. We have
\begin{equation}
\label{eq:9}
\left\{\begin{aligned}1+a-b=x+1\\
a+b=n-2.\\
\end{aligned}
\right.
\end{equation}
The solutions are
\begin{equation}
\label{eq:10}
a=\frac{n+x}{2}-1,\quad b=\frac{n-x}{2}-1.
\end{equation}
At the same time the walker in position $-1$ walks straight (along left) for $b$ steps and then turns around (along right) for $a$ steps. With the same values of $a$ and $b$ in Eq. (\ref{eq:10}), the walker starts from the position $-1$ arrives at the position $(x+1)$ at the $(n-1)$th step controlled by the coin state.

The state of the walker-coin system after the $i$th step is
\begin{align}
\label{eq:11}
|\psi\rangle^{i}=&\alpha|-i+2(i-b-1)\varepsilon(i-b-2)\rangle|\varepsilon(b+1-i)\rangle\notag\\
&+\beta|i-2(i-a-1)\varepsilon(i-a-2)\rangle|\varepsilon(i-a-2)\rangle,
\end{align}
where $\varepsilon(x)$ is unit step function, which equals to $1$ when $x\ge0$, otherwise $0$. With the values of $a$ and $b$ in Eq. (\ref{eq:10}), the final state of the system after the $n$th step is $|\psi\rangle^{n}=|x\rangle(\alpha|0\rangle+\beta|1\rangle)$. Thus the goal to transfer an unknown coin state to a certain position at arbitrary $n$ steps is successfully achieved. A simple example of transferring the coin states from position $|0\rangle$ to $|3\rangle$ via $5$ steps discrete-time QW is shown in Fig. 1 (c).

The scheme can be used as an efficient quantum routing with each achievable position corresponding to a user located on the network. It is necessary to notice that not all the positions are possible to be routed to within a fixed number of steps (restricted by Eqs. (\ref{eq:6}) and (\ref{eq:7})). However, we can let the achievable positions correspond to the users, while others act as ancillary positions.

If the walking continues with the coin flipping operator $\sigma_{x}$ (like what we do to the initial state at the $1$st step), the state of the system after the $(n+1)$th step is
  \begin{equation}
  \label{eq:12}
  |\psi\rangle^{n+1}=\beta|x+1\rangle|0\rangle+\alpha|x-1\rangle|1\rangle.
  \end{equation}
Keeping the flipping operators corresponding to the positions as before, the states of the following steps are
  \begin{align}
  \label{eq:13}
  |\psi\rangle^{n+i}=&\alpha|x-i+2(i-a-1)\varepsilon(i-a-2)\rangle|\varepsilon(a+1-i)\rangle\notag\\
                   &+\beta|x+i-2(i-b-1)\varepsilon(i-b-2)\rangle|\varepsilon(i-b-2)\rangle.
  \end{align}
The state after the $2n$th step is
  \begin{equation}
  \label{eq:14}
  |\psi\rangle^{2n}=|0\rangle(\alpha|0\rangle+\beta|1\rangle).
  \end{equation}
Thus the system interestingly back to the initial states (both position and coin state). Iterating the process shows that the process has a periodicity, which has also been seen in many important works about state transfer~\cite{Kendon}, with period of $2n$.
  \begin{figure}[htb]
  \includegraphics[width=8cm]{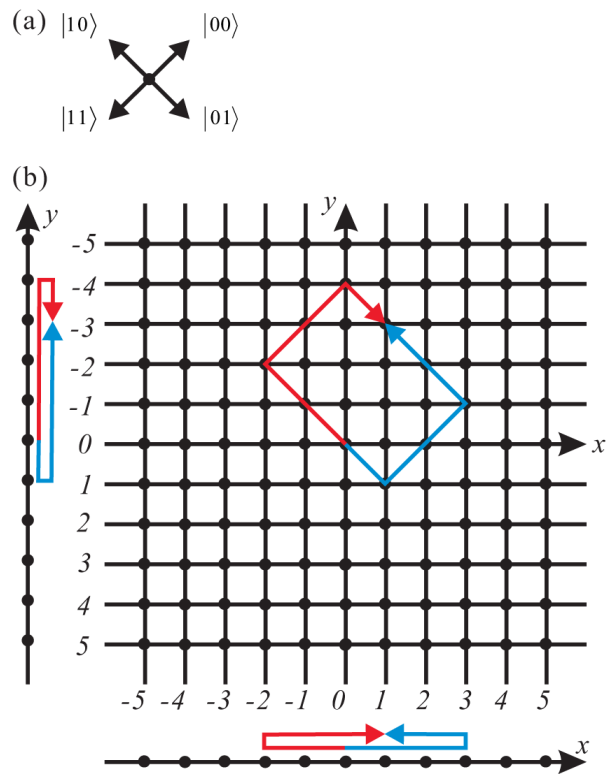}
  \caption{The scheme in 2-dimensional case is shown in the graph. (a) The walking direction correspond to the resulting coin states. (b) Perfectly transferring the coin state $\alpha|01\rangle+\beta|10\rangle$ from the original position (0,0) to the target position (1,3) after 5 steps. The first walker walks along the $x$ axis and the second walker walks along the $y$ axis. The red and blue arrows indicate the information flows of $\alpha$ and $\beta$ respectively. }
  \label{fig:2}
  \end{figure}
\subsection{Routing entangled coins in 2D discrete-time QW}
Now let us think about the 2D discrete-time QWs, in which the coins control the walker to walk different directions (seeing Fig. 2 (a)). In this subsection we focus on the case in which the two coins are entangled, which is more interesting and significant for many quantum information process.

The flipping of two coins is $C=C'\otimes C''$, where the superscripts $'$, $''$ stand for the first and second coin. It is a local operation and does not break the entanglement between the two walk-coin systems. We explicitly demonstrate how to transfer the unknown entangled coin state $(\alpha|01\rangle+\beta|10\rangle)$ from the original position $(0,0)$ to the final position $(x,y)$ after $n$ steps. The whole process can be described as
\begin{equation}
\label{eq:15}
|00\rangle^{0}(\alpha|01\rangle+\beta|10\rangle)\xlongrightarrow{\text{n\, steps}}|xy\rangle^{n}(\alpha|01\rangle+\beta|10\rangle),
\end{equation}
where the superscripts denote the step number, and we drop the superscript on the coin state for simplicity. There are four types of biased coin flipping operators used in the 2D protocol, $\mathbb{I}\otimes\mathbb{I}$, $\mathbb{I}\otimes\sigma_z$, $\sigma_{z}\otimes\mathbb{I}$, and $\sigma_{z}\otimes\sigma_{z}$. One can set the coin flipping operator $\mathbb{C}{_{nm}^{(i)}}$ of the $i$th walker in position $m$ at the $n$th step to be $\sigma_z$ when one wants the walker to turn around at position in the step, otherwise to be $\mathbb{I}$. The whole scheme is expressed as below.

Setting $\mathbb{C}'_{00}=\mathbb{C}''_{00}=\sigma_x$, i.e. $\mathbb{C}=\sigma_x\otimes\sigma_x$, the state after $1$st step is
\begin{equation}
\label{eq:16}
|\psi\rangle^1=|1,-1\rangle^1\beta|01\rangle+|-1,1\rangle^1\alpha|10\rangle.
\end{equation}
The state of the first coin is in $|0\rangle$ and $|1\rangle$, corresponding to the walker in position $1$ and $-1$. Routing it to $(x+1)$ and $(x-1)$ is actually similar to 1D case surveyed before. According to Eqs. (\ref{eq:5}) and (\ref{eq:6}), the walker in position $1$ walks straight (along right) for \begin{equation}
\label{eq:17}
a_1=\frac{(n+x)}{2}-1
\end{equation}
steps, turns around at $|a_1+1\rangle$ during the $(a_1+1)$th step, then walks straight (along left) for \begin{equation}
\label{eq:18}
b_1=\frac{(n-x)}{2}-1
\end{equation}
steps to reach $(x-1)$ at the $(n-1)$th step. At the same time, the walker in $-1$ walks straight (along left) for $b_1$ steps, then turn around at $(-b_1-1)$ during the $(b_1+1)$th step, then walks straight (along right) for $a_1$ steps will reach $|x+1\rangle$ at the $(n-1)$th step. Whereas the state of the second coin is in $|0\rangle$ and $|1\rangle$, corresponding to the walker in position $|1\rangle$ and $|-1\rangle$. Routing it to $|x+1\rangle$ and $|x-1\rangle$ is exactly same with the process of the first walker-coin system with
\begin{equation}
\label{eq:19}
a_2=\frac{(n+y)}{2}-1
 \end{equation}
 and
 \begin{equation}
 \label{eq:20}
 b_2=\frac{(n-y)}{2}-1.
 \end{equation} The four biased coin flipping operators can meet the requirements of the whole process. The following states can be similarly expressed as
\begin{widetext}
\begin{align}
\label{eq:21}
|\psi\rangle^i&=\alpha|[-i+2(i-b_1-1)\varepsilon(i-b_1-2)],[i-2(i-a_2-1)\varepsilon(i-a_2-2)]\rangle^i|\varepsilon(b_1+1-i),\varepsilon(i-a_2-2)\rangle\notag\\
&+\beta|[i-2(i-a_1-1)\varepsilon(i-a_1-2)],[-i+2(i-b_2-1)\varepsilon(i-b_2-2)]\rangle^i|\varepsilon(i-a_1-2),\varepsilon(b_2+1-i)\rangle.
\end{align}
\end{widetext}
where $\varepsilon(i)$ is the step function. With $a_i$, $b_i$ ($i=1,2$) taking the value of Eqs. (\ref{eq:17}), (\ref{eq:18}), (\ref{eq:19}) and (\ref{eq:20}), the state after the $n$th step is
\begin{equation}
\label{eq:22}
|\psi\rangle^n=|xy\rangle^n(\alpha|01\rangle+\beta|10\rangle).
\end{equation}
Thus the unknown entangled coin state is successfully routed to the target positions, in other words, successfully delivered to the users corresponding to the positions. Quantum information process based on entanglement sharing can be carried on successfully after the efficient routing. A simple example of transfer an unknown coin state $(\alpha|01\rangle+\beta|10\rangle)$ form position $(0,0)$ to $(1,3)$ is shown in Fig. 2 (b).

Continue the walking with a $\sigma_z\otimes\sigma_z$ flipping operator, and other coin flipping operators correspondent to the positions as before. We surprisingly find the coin states perfectly transferred to the initial position after $2n$ steps.

For multi coins in architecture of high dimensional (say $d$, $d\geq2$) discrete-time QW, the efficient routing scheme can be developed based on the way we develop the 1D scheme to 2D entangled case. The special setting ($\sigma_x$) of each walker-coin system is depends on the target position and the step numbers. Efficient routing $d$ coins needs $3d$ special settings. This grows linearly with the number of the coins.

\section{CONCLUSION}
In summary, we have proposed an efficient routing scheme with arbitrary unknown state transferring from input to arbitrary output ports, which is promising to be implemented in experiment. One can perfectly transferring the unknown coin state in 1D discrete-time QW from the initial position to any target position, and the settings are simple and independent of the number of the target positions. One needs set three special coin flipping operators ($\sigma_z$), while live the others to be $\mathbb{I}$. Efficient routing can be implemented based on perfect state transfer. The scheme is also efficient for routing multi-qubit entanglement. The number of special settings in routing multi-coin states in high dimensional QW grows linearly with the number of coins, which makes our protocol feasible with the current experimental technology.

Besides the potential value in quantum communication and quantum computation processes, the scheme generates periodicity in discrete-time QW on line, which is traditionally seen in discrete-time QW on periodic graph~\cite{Kendon}. This gives a deep inspiration of exploring potential application of discrete-time QW on line.
\section{ACKNOWLEDGE}
This work has been supported by NSFC under 11174052, 973 Program under 2011CB921203 and the Open Fund
from the State Key Laboratory of Precision Spectroscopy of East China Normal University.


\begin{thebibliography}{99}
\bibitem{Bose}S. Bose, Phys. Rev. Lett. {\bf 91}, 20 (2003).
\bibitem{DTQWST}P. Kurzy\'{n}ski, and A. W\'{o}jcik Phys. Rev. A {\bf83}, 062315 (2011).
\bibitem{networkST}A. W\'{o}jcik, T. {\L}uczak, P. Kurzy\'{n}ski, A. Grudka, T. Gdala, and M. Bednarska, Phys. Rev. A {\bf75}, 022330 (2007)
\bibitem{Christandl}M. Christandl, N. Datta, A. Ekert, and A. J. Landahl, Phys. Rev. Lett. {\bf92}, 18 (2004).
\bibitem{Kendon}V. M. Kendon, and C. Tamon, J. Comput. Theor. Nanosci. {\bf 8}, 422-433 (2011).
\bibitem{Chudzicki}C. Chudzicki, and F. W. Strauch, Phys. Rev. Lett. {\bf 105}, 260501 (2010).
\bibitem{QW}Y. Aharonov, Y. Davidovich, and N. Zagury, Phys. Rev. A {\bf48}, 2 (1993).
\bibitem{Kempe}J. Kempe, Contemp. Phys. {\bf44}, 4 (2003).
\bibitem{Kitagawa}T. Kitagawa, M. S. Rudner, E. Berg, and E. Demler, Phys. Rev. A {\bf82}, 033429 (2010).
\bibitem{X1}P. Xue, H. Qin, B. Tang, and B. C. Sanders, New J. Phys. {\bf16}, 053009 (2014).
\bibitem{X2}P. Xue, H. Qin, and B. Tang, Sci. Rep. {\bf4}, 4825 (2014).
\bibitem{np}A. Crespi, R. Osellame, R. Ramponi, V. Giovannetti, R. Fazio, L. Sansoni, F. D. Nicola, F. Sciarrino, and P. Mataloni, Nature photon. {\bf7}, 322-328 (2013).
\bibitem{Oliveira}A. C. Oliveira, R. Portugal, and R. Donangelo, Phys. Rev. A {\bf74}, 012312 (2006).
\bibitem{burnett}O. Buerschaper, and K. Burnett, e-print arXiv:quant-ph/0406039.
\bibitem{segwa}E. Segawa, J. Comput. Theor. Nanosci. {\bf10}, 1583-1590 (2013)
\bibitem{Childs}A. M. Childs, Phys. Rev. Lett. {\bf102}, 180501 (2009).
\bibitem{Childs2}A. M. Childs, D. Gosset, and Z. Webb, Science {\bf339}, 791 (2013)
\bibitem{Lovett}N. B. Lovett, S. Cooper, M. Everitt, M. Trevers, and V. M. Kendon, Phys. Rev. A {\bf 81}, 042330 (2010).
\bibitem{Travaglione}B. C. Travaglione, and G. J. Milburn, Phys. Rev. A {\bf65}, 032310 (2002).
\bibitem{Sanders}B. C. Sanders, S. D. Bartlett, B. Tregenna, and P. L. Knight, Phys. Rev. A {\bf67}, 042305 (2003).
\bibitem{Dur}W. D\"{u}r, R. Raussendorf, V. M. Kendon, and H. J. Briegel, Phys. Rev. A {\bf66}, 052319 (2002).
\bibitem{X09}P. Xue, B. C. Sanders, and D. Leibfried, Phys. Rev. Lett. {\bf103}, 183602 (2009).
\bibitem{X5}P. Xue, B. C. Sanders, A. Blais, and K. Lalumi\'{e}re, Phys. Rev. A {\bf78}, 042334 (2008).
\bibitem{Du}J. Du, H. Li, X. Xu, M. Shi, J. Wu, X Zhou, and R. Han, Phys. Rev. A {\bf67}, 042316 (2003).
\bibitem{Perets}H. B. Perets, Y. Lahini, F. Pozzi, M. Sorel, R. Morandotti, and Y. Silberberg, Phys. Rev. Lett {\bf100}, 170506 (2008).
\bibitem{Karski}M. Karski, L. F\"{o}rster, J. M. Choi, A. Steffen, W. Alt, D. Meschede, and A. Widera, Science {\bf325}, 174 (2009).
\bibitem{Schreiber}A. Schreiber, A. G\'{a}bris, P. P. Rohde, K. Laiho, M. \v{S}tefa\v{n}\'{a}k, V. Poto\v{c}ek, C. Hamilton, I. Jex, and C. Silberhorn, Science {\bf336}, 55 (2012).
\bibitem{Bouwmeester}D. Bouwmeester, I. Marzoli, G. P. Karman, W. Schleich, and J. P. Woerdman, Phys. Rev. A {\bf61}, 013410 (1999).
\bibitem{Zahringer}F. Z\"{a}hringer, G. Kirchmair, R. Gerritsma, E. Solano, R. Blatt, and C. F. Ross, Phys. Rev. Lett. {\bf104}, 100503 (2010).
\bibitem{Schmitz}H. Schmitz, R. Matjeschk, Ch. Schneider, J. Glueckert, M. Enderlein, T. Huber, and T. Schaetz, Phys. Rev. Lett. {\bf103}, 090504 (2009).
\bibitem{Cote}R. C\^{o}t\'{e}, A. Russell, E. E. Eyler, and P. L. Gould, New J. Phys. {\bf8}, 156 (2006).
\bibitem{Do}B. Do, M. L. Stohle, S. Balasubramanian, D. S. Elliott, C. Eash, E. Fischbach, M. A. Fischbach, A. Mills, and B. Zwickl, J. Opt. Soc. Am. B {\bf22}, 499 (2005).
\bibitem{Zhang}P. Zhang, X. F. Ren, X. B. Zou, B. H. Liu, Y. F. Huang, and G. C. Guo, Phys. Rev. A {\bf75}, 052310 (2007).
\bibitem{Peruzzo}A. Peruzzo, M. Lobino, J. C. F. Matthews, N. Matsuda, A.Politi, K. Poulios, X. Q. Zhou, Y. Lahini, N. Ismail, K. W\"{o}rhoff, Y. Bromberg, Y. Silberberg, M. G. Thompson, and J. L. OBrien, Science {\bf329}, 1500 (2010).
\bibitem{Broome}M. A. Broome, A. Fedrizzi, B. P. Lanyon, I. Kassal, A. Aspuru-Guzik, and A. G. White, Phys. Rev. Lett. {\bf104}, 153602 (2010).
\bibitem{Schreiber2}A. Schreiber, K. N. Cassemiro, V. Poto\v{c}ek, A. G\'{a}bris, P. J. Mosley, E. Andersson, I. Jex, and Ch. Silberhorn, Phys. Rev. Lett. {\bf104}, 050502 (2010).
\bibitem{Sansoni}L. Sansoni, F. Sciarrino, G. Vallone, P. Mataloni, A. Crespi, R. Ramponi, and R. Osellame, Phys. Rev. Lett. {\bf108}, 010502 (2012).
\bibitem{teleportation}C. H. Bennett, G. Brassard, C. Cr\'{e}peau, R. Jozsa, A. Peres, and W. K. Wootters, Phys. Rev. Lett. {\bf 28} 1895 (1993).
\bibitem{Brun}T. Brun, I. Devetak, and M. H. Hsieh, Science {\bf314}, 436 (2006).
\bibitem{Jozsa}R. Jozsa, and N. Linden, Proc. R. Soc. London Ser. A {\bf459}, 2011 (2003).
\bibitem{line}A. Nayak, and A. Vishwanath, e-print arXiv:quant-ph/0010117.
\end{thebibliography}
\end{document}